\documentclass[12pt,twocolumn]{article}
\usepackage{subeqn}
\title{Solving large linear algebraic systems
in the context of integrable non-abelian Laurent ODEs}
\author{
Thomas Wolf, Brock University, Ontario, Canada\\
email: twolf@brocku.ca\vspace{12pt}\\
Eberhard Schr\"{u}fer, Bonn, Germany\\
email: eberhard.schruefer@ca-musings.de\vspace{12pt}\\
Kenneth Webster, Brock University, Ontario, Canada\\
email: kw07ty@brocku.ca
}

\begin{document}

\maketitle
\begin{abstract}
  The paper reports on a computer algebra program {\sc LSSS} (Linear Selective
  Systems Solver) for solving linear algebraic systems with rational
  coefficients. The program is especially efficient for very large sparse
  systems that have a solution in which many variables take the value
  zero. The program is applied to the symmetry investigation of a non-abelian
  Laurent ODE introduced recently by M.\ Kontsevich. The computed symmetries
  confirmed that a Lax pair found for this system earlier generates all first
  integrals of degree at least up to 14.
\end{abstract}

\section{Introduction}
In mathematics, science and engineering many problems lead to sparse
linear algebraic systems. For example, in a discretization of a smooth
object the relations between variables defined in neighbouring points
typically involve only a small number of variables that is dependent
on the dimension of the object which is low.

For sparse linear systems there are computer algebra programs available,
for example, the code of Roman Pearce \cite{RoPe} that has been
incorporated into {\sc Maple} 14: {\tt SolveTools:-Linear}. It is
automatically applied if a sparse system is to be solved.

The typical concern of solvers for sparse systems is to avoid a choice
of pivot that increases the number of variables in equations and thus
leads to non-sparse equations during the solution process. For many
problems this can not be avoided, only delayed. For example, when
solving the sparse linear system resulting from the discretization of
the (partial differential) Laplace equation (i.e.\ heat equation in
the stationary limit) it is clear that the temperature at each single
point depends on the temperature at {\em all} boundary points and that
the temperature inside is typically non-zero.\footnote{Even if the
  temperature on the whole boundary is zero except on an
  $\varepsilon$-sized part where it is positive, the temperature at
  all inner points will be positive.}

The linear systems we studied are also sparse but different in nature.
The value of most of their variables in the solution is exactly zero. 
Such systems play the role of selectors and are therefore called ``selection
systems'' in this paper, because from a large number of monomials, all
with a constant undetermined coefficient, some linear combinations of the
monomials are selected which satisfy some condition and the other monomials
are dropped by setting their coefficient to zero. This is very different from
linear sparse systems like those resulting from the discretization of the
Laplace equation. Table \ref{table0} compares both types of problems.

\begin{table*}[ht]   
   \begin{center}   
   \begin{tabular}{|l|c|c|} \hline
    type    & ``numerical'' systems    & ``selection'' systems          \\ \hline
    examples& systems resulting from a & systems resulting from a sym-  \\
            & discretization of PDEs   & metry investigation of PDEs    \\ \hline
    value of free parameters   & any floating     & 0 or 1              \\
    when applying the solution & point numbers    & (to isolate the individual  \\ 
    of the linear system       & (boundary values of PDE) & symmetries) \\ \hline
    number of zero-valued      & essentially none & most variables \\
    variables in solution      &                  &                \\ \hline
    sparsity                   &      yes         &     yes        \\ \hline 
    overdetermination          &       no         &     yes        \\ \hline 
    usability of iteration     &                  &                \\
    schemes for large problems &   useful         & not useful     \\
    of that type               &                  &                \\ \hline 
   \end{tabular}
   \end{center}  
   \caption{Characterization of two different types of sparse linear systems}
  \label{table0}
\end{table*} 

Selection systems have a number of useful properties.
\begin{itemize} \itemsep=0pt
\item The existence of zero-valued variables allows an effective solution as
  performed by the program {\sc LSSS}, described in this paper. {\sc LSSS} is 
  running under the computer algebra system {\sc Reduce} (\cite{RedSource}).
\item From the application it is clear whether a linear system belongs to this
  class, i.e.\ whether many variables take the value zero and thus 
  a specialized computer program should be applied.
\item The efficiency of solving selection systems increases with the
  complexity of the problem. For example, the higher the degree of the
  polynomial ansatz in a symmetry investigation - the larger is the size of
  the linear system for the unknown coefficients but the higher is also the
  percentage of zero-valued coefficients and the higher is the efficiency of
  the solution technique that uses zero-valued variables (see section
  \ref{charac}).
\item Not only can such linear systems be solved efficiently, they can also be
  formulated more economically. More precisely, it is possible to formulate
  much smaller equivalent systems as shown in section \ref{iter}.
\end{itemize}

The strategy to have many different methods available to solve a system of
equations and to give those methods the highest priority which make most
progress and are least risky to lead to an explosion of size is not new.  In
the {\sc Maple} command {\tt solve} this is used since 1985 even with the
heuristic of substituting zero variables first (see \cite{GoMo}) and in the
{\sc Reduce} package {\sc Crack} such strategies including substitution of
variables by zero as highest priority are applied to the solution of
overdetermined algebraic and differential systems (\cite{Wo04}. What is new in
this paper is the realization that if variables take the value of zero in
linear systems, then usually many variables vanish and then
very fast routines for identifying 1-term equations, for dropping such
variables from equations and even for avoiding the construction of equations
pays of.

What justifies the creation of special purpose programs for selection
systems is the fact that these systems occur frequently, especially in
integrability investigations of differential equations when computing
infinitesimal symmetries, first integrals, conservation laws,
variational principles or other qualitative properties which may but
need not exist and for which an ansatz can be formulated.

The following section describes an application that leads to a series
of sparse systems with a high percentage of zero-valued variables.
This application was the starting point for the development of {\sc
  LSSS}. Readers interested essentially in the computational aspects
of formulating and solving the systems may proceed directly to section
\ref{charac} which describes methods for formulating and solving
selection systems efficiently.

The times for solving the linear systems resulting in the applications
are shown in section \ref{results}.  The subsequent section
\ref{reduce} discusses aspects of the computer algebra system {\sc
  Reduce} which become important in large computations. Sections
\ref{maple} and \ref{linbox} report on tests of other computer algebra
systems. The paper concludes with a list of available procedures in
section \ref{procedures} and a summary.

\section{The application} \label{appli}

\subsection{A non-abelian ODE-system}

In the programme of M.\ Kontsevich of creating a proper
non-commutative algebraic geometry inspired by modern quantum
field-theoretic requirements and challenges, he proposed a
non-commutative version of symplectic geometry in which integrable
non-abelian ODE-systems appear naturally (\cite{MKpriv}).

More specifically, he considered the discrete map
\begin{equation} \label{dsym}
  u \ \rightarrow \ u v u^{-1} \ , \ \ \ \
  v \ \rightarrow \ u^{-1} + v^{-1} u^{-1}
\end{equation}
(\cite{MKtalk}) and the following non-abelian ODE system
which is invariant under (\ref{dsym}):
\begin{equation}\label{inhom}
u_t=u v -u v^{-1} - v^{-1}, \ v_t=- v u + v u^{-1} + u^{-1}
\end{equation}
where $u,v$ are non-commutative variables (in particular, square
matrices of arbitrary size). In comparison to non-abelian ODE
systems with polynomial right hand sides investigated first in
\cite{olsok} and later in \cite{MikSok}, \cite{Ef} the system
(\ref{inhom}) involves Laurent polynomials, i.e.\ polynomials in $u,v$
and inverses $u^{-1}, v^{-1}$.

Based on numerical experiments Kontsevich conjectured that
(\ref{inhom}) is integrable itself. This was demonstrated recently in
\cite{WolEfi11} where a Lax pair has been found which also proves
integrability of (\ref{dsym}) (see section 2 of \cite{MKtalk}). In
addition a pre-Hamiltonian operator is given in \cite{WolEfi11} that
maps gradients of trace first integrals to infinitesimal symmetries.
The existence of infinitely many symmetries and even better of a Lax
pair is taken as an indicator of integrability of an ODE or PDE system
(\cite{olver}). To verify that the Lax pair produces all Laurent
polynomial first integrals our strategy is to compute by brute force
all Laurent polynomial infinitesimal symmetries up to some degree and
to compare them with the symmetries produced from the Lax pair. This
paper reports on the computer algebra program {\sc LSSS} that was
developed to solve the linear algebraic systems resulting in the
computation of all such symmetries up to degree 16.

\subsection{Symmetries} \label{sy}
For a given system of equations
\begin{equation}
\label{eq2}
u_{t}=P_{1}(u,v,u^{-1},v^{-1}), \ \ \ v_{t}=P_{2}(u,v,u^{-1},v^{-1})
\end{equation}
where $P_1, P_2$ are polynomials in $u,v,u^{-1},v^{-1}$ a Laurent
polynomial (infinitesimal) symmetry is defined through two polynomials
$Q_1, Q_2$ in $u,v,u^{-1},v^{-1}$ such that the flow
\begin{equation}
\label{symm}
u_{\tau}=Q_{1}(u,v,u^{-1},v^{-1}), \ \ v_{\tau}=Q_{2}(u,v,u^{-1},v^{-1})
\end{equation}
commutes with the system, i.e.\ 
\begin{subequations}  \label{symcon}
 \begin{equation} 
   D_\tau D_t \ u = D_t D_\tau \ u,  \label{symcon1} 
 \end{equation}
 \begin{equation} 
   D_\tau D_t \ v = D_t D_\tau \ v . \label{symcon2}
 \end{equation}
\end{subequations}
The vector $(Q_{1},Q_{2})^{t}$ is called generator of the symmetry.

In this paper the system (\ref{eq2}) is given through (\ref{inhom}).
The polynomials $Q_1, Q_2$ are generated by a special purpose procedure that
creates the most general (inhomogeneous) polynomials up to some given
degree in the non-abelian variables $u,v,u^{-1},v^{-1}$ with symbolic
(abelian) coefficients $c_i$. The only condition satisfied by
$Q_1,Q_2$ is that $u,u^{-1}$ and $v,v^{-1}$ are not standing next to
each other anywhere in any term as they would cancel. 

The symmetry conditions (\ref{symcon}) have to be satisfied identically for 
any $u, v$. That means, after (\ref{symcon}) is formulated, the coefficients
of all different products of powers of $u,v,u^{-1},v^{-1}$ have to be set to
zero generating a linear system for the unknown coefficients in the 
symmetry generators(\ref{symm}). This process is called 'splitting' in the
remainder of the paper.

The number $\hat{t}_n$ of terms of each $Q_i$ of degree $n$ (which is one half
of the number of unknown coefficients) satisfies the recursive relation
$\hat{t}_n = 3\hat{t}_{n-1}+2$ with $\hat{t}_0=1$ because apart from the
coefficients $c_i$ the polynomial of degree 0 has only the term 1 and the most
general polynomial of degree $n$ is obtained by multiplying each term of the
most general polynomial of degree $n-1$ from one side, say from the left, with
all possible three of the four factors $u,v,u^{-1},v^{-1}$ which do not give a
cancellation. One exception is the term 1 that is multiplied with each one of
the 4 factors i.e.\ from this term results one extra term and the extra term 1
is added giving $\hat{t}_n = 3\hat{t}_{n-1}+1+1 = 3\hat{t}_{n-1}+2$.  Thus the
total number of terms $t_n$ occuring in $Q_1$ and $Q_2$ of degree $n$
satisfies $t_n = 3t_{n-1}+4$.

\subsection{Necessary symmetry conditions}
Additional information available for our symmetry computation results
from a separate computation of full first integrals $I$ satisfying
$D_t I = 0$ where $D_t$ is the total time-derivative. The condition
that a Laurent polynomial in $u,v$ is a first integral is very
restrictive because it implies that each of the $m\times m$ components
of the matrix first integral is an integral of its own. Restrictive
conditions lead to very overdetermined systems (here linear) which are 
easier to solve. Therefore it was possible to compute all first integrals 
up to degree 14 with the result that 
\begin{equation} \label{FI}
 I = u v u^{-1} v^{-1},\ \ \ I^{-1} = v u v^{-1} u^{-1}
\end{equation}
and their integer powers $I^k, \ \ k=-3,..,3$ are the only first
integrals up to degree 14.

 From the symmetry conditions (\ref{symcon})
and $D_t I = 0$ follows
\[ D_tD_\tau I=D_\tau D_t I = 0\]
and further from $I^k$ being the only first integrals (up to degree 14)
\begin{subequations} \label{nc}
\begin{equation} \label{nc1}
 D_\tau I = \sum_{k=-k_0}^{k_0} a_k I^k, 
\end{equation}
and similarly
\begin{equation} \label{nc2}
 D_\tau I^{-1} = \sum_{k=-k_0}^{k_0} b_k I^k, 
\end{equation}
\end{subequations}
for sufficiently high $k_0$, in our case $k_0=3$ because $I$ is of degree 4. 
These conditions involve a few more extra unknown constants $a_k, b_k$ 
but adding (\ref{nc2}) is beneficial because 
these are first order conditions involving fewer terms than the
full symmetry conditions (\ref{symcon}) which are of second order.

\section{The solution of selection systems} \label{charac}

The linear systems resulting from splitting the necessary and
sufficient conditions (\ref{symcon}) and the additional necessary
conditions (\ref{nc}) cover a wide range of sizes as shown in table
\ref{table1}. As discussed in section \ref{sy}
the numbers $k_n$ of unknowns for a symmetry ansatz of degree $n$
grow like $k_{n+1}=3k_n+4$ which is also the growth rate of the number
$e_1$ of equations in the necessary condition $D_\tau I=0$.\footnote{The
   vanishing of constants $a_k$ in (\ref{nc1}) and $b_k$ in (\ref{nc2})
   follows when formulating these conditions.}

\begin{table*}[ht]
  \tabcolsep=3pt
  \begin{center}
  \begin{tabular}{|r|r|r|r|r|r|r|} \hline
$n$ &  $k$       & $e_1$      & $t_1$      & $e_2$       & $t_2$        & $p$ \\ \hline
 3  & 106        & 142        & 192        & 448         & 1,034        &  1 \\
 4  & 322        & 430        & 616        & 1,412       & 3,706        &  2 \\
 5  & 970        & 1,294      & 1,904      & 4,448       & 12,914       &  4 \\
 6  & 2,914      & 3,886      & 5,784      & 13,878      & 44,098       &  5 \\
 7  & 8,746      & 11,662     & 17,440     & 43,052      & 148,346      &  7 \\
 8  & 26,242     & 34,990     & 52,424     & 132,954     & 493,162      &  8 \\
 9  & 78,730     & 104,974    & 157,392    & 409,470     & 1,623,842    & 12 \\
 10 & 236,194    & 314,926    & 472,312    & 1,258,526   & 5,304,562    & 13 \\
 11 & 708,586    & 944,782    & 1,417,088  & 3,862,086   & 17,212,778   & 17 \\
 12 & 2,125,762  & 2,834,350  & 4,251,432  & 11,835,758  & 55,535,578   & 18 \\
 13 & 6,377,290  & 8,503,054  & 12,754,480 & 36,228,892  & 178,298,450  & 24 \\
 14 & 19,131,874 & 25,509,166 & 38,263,640 & 110,777,292 & 569,970,466  & 25 \\ 
 15 & 57,395,626 & 76,527,502 & $>1.1\!\times\!10^8$ & $>3.3\times 10^8$ & $> 1.7\times 10^9$ & 31 \\ 
 16 & $>172\!\times\!10^6$ & 
      $>229\!\times\!10^6$ & 
      $>3.4\!\times\!10^8$ & 
      $>1\!\times\!10^9$   &
      $>5.7\!\times\!10^9$ & 32 \\ \hline

  \end{tabular}   
  \end{center}
  \caption{
 For each symmetry ansatz of degree $n=3..16$ are listed the numbers: 
 $k$ of variables, 
 $e_1$ of equations and $t_1$ of terms of system $D_\tau I=0$,
 $e_2$ of equations and $t_2$ of terms of system $[D_t,D_\tau](u,v)=0$ and
 $p$ of free parameters of the solution.}
 \label{table1}
\end{table*}

As the table indicates, the linear systems have a few characteristic
properties that need to be exploited in order to compute high degree
symmetries. These properties are:
\begin{itemize}
\item {\em Overdetermination:} There are 4.5 to 5.5 times as 
      many equations as unknowns. 
\item {\em Sparsity:} Equations have on average 4 to 5 terms. 
\item {\em Zero-valued Variables:} Most of the variables take an exact value of
      zero in the solution. Even more importantly, the percentage of zero-valued 
      variables is increasing as the degree of the ansatz increases, i.e.\ as 
      the system to be solved increases in size. For systems resulting from symmetries 
      of degree 4 this percentage is 93.2\% and for symmetries of degree 16 the
      percentage is 99.966\% . Nevertheless, the solution for degree 16 is not trivial.
      It has 58118 non-vanishing variables and 32 free parameters.
\end{itemize}

\subsection{Ideas for a solver of large sparse linear algebraic systems} \label{ideas}
The following ideas take advantage of the features of selection
systems as listed above and describe the stages of development 
of the computer algebra program {\sc LSSS}. These stages have been to:
\begin{itemize}
\item start with a program for solving a stream of equations,
\item sort equations initially by size, shortest first, 
\item apply 1-term equations very efficiently and apply them
      before any other equation,
\item be able to generate only 1-term equations before formulating the
      whole system,
\item iterate between the exclusive formulation and the application 
      of 1-term equations,
\item be able to generate and take advantage of extra 
      necessary conditions before working on the original system, and
\item choose from possibly different options to generate 1-term 
      equations the most efficient one.
\end{itemize}
Any extra low level procedures that were written to implement
these concepts assume that systems are linear in the unknowns.
This allows them to be more efficient than universal routines.

In the following subsections more detailed comments are made to each 
of the measures.

\subsection{A stream of equations} \label{stream}
In earlier work \cite{TsaWo} large and very overdetermined polynomial
systems were solved which had too many equations (in the order of
$10^{14}$) to be even formulated initially. A linear algebraic system
solver was developed at the time for related problems dealing with the
system of equations as an incoming stream of data not to be stored in
RAM memory, only to be processed equation by equation. A strength of
this algorithm is that any limitation on available RAM memory poses
only restrictions on the size of the preliminary solution of the
system and thus only on the number of variables, not on the number of
equations. 

At the start of this program a preliminary solution is initialized as
an empty list. Then with each incoming equation the following steps are performed.
\begin{itemize}
\item The equation is simplified modulo a preliminary
  solution stored in the form of a substitution list $g_i = f_i(g_j)$
  where $g_k$ are the unknowns and $f_i$ are linear expressions in the unknowns
  except $g_k$ from any of the left hand sides.
\item If the resulting simplified equation  is not an identity then it 
  is solved for one of the $g_m$, 
\item substitutions $g_m=f_m(g_j)$ are performed in all $f_i$ of the
  preliminary solution, and
\item $g_m=f_m(g_j)$ is appended to the substitution list.
\end{itemize}

In the computation of high order symmetries this procedure is applied to the 
remaining system after all 1-term equations have been determined and used.
The remaining system is small enough that it does not have to be stored
on hard disk and then read again from disk. 

\subsection{Sorting equations} \label{sort}
A first speedup is obtained by sorting equations according to size,
shortest first, before processing them as described above leading to
the times shown in column (B) of table \ref{table2}.  For example,
for symmetries of degree 9 the speed up is a factor of 4.

Processing shorter equations earlier means that the substitution list
involves shorter right hand sides and reduces incoming equations to
shorter size earlier which adds shorter new substitution rules to the
preliminary solution, i.e.\ it is a self-amplifying increase of efficiency.

In symbolic mode of {\sc Reduce} the sorting of equations can be done very
effectively, for example, by establishing a list $L=\{l_1,l_2,l_3,...\}$ where $l_i$
is a list of (pointers to) all equations with $i$ terms and then
linking these lists. 
Assigning an equation to a list $l_i$ is done efficiently by having 2 pointers, one
stepping from one term to the next in the equation and the other at
the same time stepping from $l_i$ to $l_{i+1}$.  When the first
pointer reached the end of the equation, the other pointer gives
the list $l_i$ under which the equation is then listed.

\subsection{Applying 1-term equations} \label{apply}
Sorting equations is beneficial if equations vary much in size. This
is especially the case for selection systems 
with many equations having only one term.

But not all the variables that take zero value in the solution need to
appear in 1-term equations in the original system. Many 1-term
equations may result only after the first 1-term equations are
applied. Still, during the solution process many 1-term equations are
generated which justifies their special treatment.

What makes 1-term equations special is the fact that replacing a
variable by zero in a polynomial (or linear) expression can be done
very fast without re-writing the expression, at least in a
lisp-representation: the pointer from the previous term to the term
that vanishes is simply re-directed to the next term. Also, no
re-simplification is necessary. In contrast, already the
application of 2-term equations requires afterwards 
simplifications as the term resulting from
substitution may combine with other terms.

To set many variables in an expression to zero in a most efficient way, the
value cell of the variable was set to nil. Testing a variable for a zero value
can then be done by just testing whether the variable is bound.  After this is
done for all known vanishing variables, the (large linear) expression in these
variables is pruned only once by the special low level routine {\tt
  PruneZeros}. The pruning is done 'in place' avoiding copying of the whole
expression and also reducing future garbage collections.

All efficiency improving measures introduced in the sections
\ref{stream}, \ref{sort}, \ref{apply} are applied in the procedure
{\sc LSSS} \cite{liso} but can also be performed alone on given
expressions.

In the following subsections techniques are described that make the
formulation of the linear system more efficient, or avoid the 
formulation of large selection systems.

\subsection{Selective splitting of equations} \label{selective}
The following steps are all computationally expensive: 
\begin{itemize} \itemsep=0pt
\item the formulation of two large symmetry conditions (\ref{symcon}), 
\item their separation (called 'splitting' in the following) into two large
  linear systems with many redundant equations by setting coefficients of
  different products of powers of $u, v$ separately to zero, and
\item millions of simplifications of the large linear systems due to 
millions of vanishing variables. 
\end{itemize}
The idea is to avoid most of these computations by
\begin{itemize} \itemsep=0pt
\item extracting selectively only 1-term equations from the
symmetry conditions (\ref{symcon}), and 
\item using them to simplify the symmetry conditions themselves, and 
\item to repeat this procedure as long as 1-term equations can be found (see section \ref{iter}).
\end{itemize}
Finally, the much smaller symmetry conditions are completely split and
the resulting linear system is solved using {\sc LSSS} as
described in the previous three subsections.

The only remaining large step is the initial formulation of the symmetry
conditions. One can save half of that computation by formulating only
one symmetry condition, extracting from it and applying to it
repeatedly 1-term equations, then pruning the ansatz for the symmetry
(\ref{symm}) before computing the second symmetry condition.

Furthermore, even the formulation of the first symmetry condition can
be postponed if short additional necessary conditions are known as
described in the next subsection. From them 1-term equations, i.e.\
vanishing variables can be extracted and the ansatz for the symmetry 
(\ref{symm}) be pruned before any new symmetry condition is formulated.

The interplay between formulating and solving equations may be the
only way to approach a large problem which otherwise could even not be
formulated.  For example, this iterative approach was necessary to
compute symmetries of degree 14 on the available computer hardware
with 128 GB memory under PSL {\sc Reduce}. The same approach was also
applied in \cite{TsaWo} to solve a non-linear system which was too
large to be formulated at once.

The key to be able to selectively split up 1-term equations is a low level
procedure written in {\sc Symbolic Mode of Reduce} that takes only two lines.
It recursively steps through an expression and identifies 1-term coefficients
of any monomial in $u,v,u^{-1},v^{-1}$. If such a coefficient is found, it
directly sets the value cell of the variable to nil (see section \ref{apply}
for the case that a system of linear equations is given from which 1-term
equations are picked).

\subsection{Utilizing additional necessary conditions}
Apart from a system of equations to be solved sometimes additional
necessary conditions are available. The question is whether these
conditions can be utilized to speed up the solution of the original
system. If a program is performing its computation with the whole
system that is to be solved at once then solving in addition extra
conditions results in an increase of the total computation time. On
the other hand, if the program is able to extract information from the
system selectively then having extra necessary equations available
means that the program has more options and may be able to solve the
combined system faster than the original system alone.

If a program is able to generate repeatedly 1-term equations and to
apply them then it is beneficial for the program to have extra
necessary conditions (\ref{nc1}). Even if such opportunities do not
exist, extra necessary conditions can already be useful if the length
of equations varies much and if all equations are sorted by size and
if system plus extra necessary conditions are very overdetermined.

\subsection{Optimizing iterations} \label{iter}
An optimal strategy to speed up computation is not to find and use as many as
possible vanishing variables but to find and use as many as possible {\em per
time}. This means one wants to find an optimum between formulating new
conditions (\ref{nc}), (\ref{symcon})
which each provide many vanishing variables but which take
considerable time to formulate, or to utilize already formulated
conditions by extracting again vanishing variables, utilizing them, extracting
more, and so on. The second process is faster than formulating new
conditions but the number of vanishing variables that are found is gradually
decreasing as shown in table \ref{table10}. 
The following comments refer to the computation of symmetries of
degree 13.\footnote{Given times refer to the computation of degree 13
  symmetries on a single CPU of a 8 core node (2 sockets x 4 cores per
  socket) Xeon @ 2.93 GHz using 48 GB memory of the node.}

After extracting vanishing variables from the necessary condition (\ref{nc1}) 
one has the options to 
  \begin{itemize} \itemsep=0pt
  \item prune $D_\tau (u,v)$ (8 sec) and re-formulate (\ref{nc1}) (39 sec), or
  \item prune $D_\tau (u,v)$ (8 sec) and formulate the other necessary condition 
  (\ref{nc2}) (39 sec)
  \item prune the already formulated condition (\ref{nc1}) (37.3 sec) \vspace{-3pt}
  \end{itemize}
  before extracting vanishing variables from that equation which was
  just formulated or pruned. Because in all 3 cases about 1,535,270
  new vanishing variables are found, the third option is the fastest
  and most efficient one which is therefore repeated several times.
\begin{table*}[ht]
  \tabcolsep=3pt
  \begin{center}
  \begin{tabular}{|r|r|r|r|r|r|r|r|r|} \hline
runs             &   1   &   2   &  3   &  4  &  5 & 6  & 7  & 8   \\ \hline
$t_1$            &  116  &  37.3 & 14.6 & 12.2&11.7&11.7&11.7&11.7 \\
$t_2$            &  1.8  &  .49  &  .32 & .26 & .26&.25 &.25 &.25  \\
new              &5314423&1535271&288684&32076&3564&396 &44  & 0   \\ \hline
  \end{tabular}
  \end{center}
  \caption{Times and results for extracting 1-term equations repeatedly
in the computation of symmetries of degree 13.
$t_1$: time to formulate (\ref{nc1}) in run 1 and to prune it in later runs,
$t_2$: time to 1-term-split (\ref{nc1}), 
new: number of new vanishing variables found in each run.}
 \label{table10}
\end{table*}
As shown on table \ref{table10} the number of vanishing variables that
are found, shrinks and the time to scan the condition stays constant,
thus it pays off after three runs to invest in the formulation of a
new condition
\begin{equation}
u_{\tau t} = u_{t \tau}.  \label{symcon3} 
\end{equation}
After a first selective split of (\ref{symcon3}) it is most effective
to prune and selectively split (\ref{nc1}) again, even twice before
continuing to prune and split selectively (\ref{symcon3}). It turns
out that formulating in addition the second symmetry condition
\begin{equation}
v_{\tau t} = v_{t \tau}  \label{symcon4} 
\end{equation}
only for the purpose of selective splitting does not result in
additional 1-term equations and is therefore not beneficial.

To summarize, denoting the pruning and selective splitting of
necessary condition (\ref{nc1}) by $n$ and the pruning and selective
splitting of symmetry condition (\ref{symcon3}) by $s$, the sequence
$n^3(snn)^4(sn)^4$ followed by a formulation of the symmetry condition
(\ref{symcon3}), a complete splitting of (\ref{symcon4}) and complete
splitting of what is left of (\ref{symcon3}) and a call of {\sc
  LSSS} to solve the linear system brings down the total time for
formulating and solving symmetry conditions for degree 13 to 467 sec,
compared to 3615 sec when straightforwardly formulating and splitting
(\ref{symcon3}), (\ref{symcon4}) and solving them with {\sc LSSS}.

An additional benefit of utilizing 1-term equations consists in a
drastic reduction of memory needs to only 13 GB for PSL {\sc Reduce}
or 7 GB for CSL {\sc Reduce} (see section \ref{reduce} about the
differences between PSL and CSL {\sc Reduce}) whereas the ad hoc formulation
of the large symmetry conditions (\ref{symcon3}), (\ref{symcon4})
requires 120 GB in PSL {\sc Reduce} or 60 GB in CSL {\sc Reduce} for
degree 13 symmetries.  Without selective splitting of 1-term
equations, i.e.\ determination of vanishing variables, it would not
have been possible to determine symmetries of degree 15 on the
computer with 128 GB and of degree 16 on a computer with
256 GB.

\subsection{Applying Solutions of large linear Systems}
In applications the solution of a system of equations has usually to be
substituted in expressions that are the main interest of the application. But
millions of variables to be substituted in expressions with millions of terms
is expensive. In such situations the labelling of zero-valued variables and
the pruning of large expressions containing them becomes useful again.

In table \ref{table2}, column (D) subst.\ shows drastically reduced
substitution times of the solution into the symmetry ansatz (\ref{symm})
compared to column (C) subst. After the computation in column (D) no explicit
substitutions are necessary. Any expression containing the unknowns needs
only a) to be pruned to drop terms involving zero-valued variables and b) be
simplified where non-zero variables get replaced by their computed value.

\section{Results} \label{results}

\begin{table*}[ht]
  \begin{center}
  \tabcolsep=3pt
  \begin{tabular}{|r|r|r|r|r|r|r|r|r|} \hline 
  &\multicolumn{2}{c|}{(A):}      
  &\multicolumn{2}{c|}{(B):}      
  &\multicolumn{2}{c|}{(C): 1-term equ.,} 
  &\multicolumn{2}{c|}{(D):} \\
n &\multicolumn{2}{c|}{}            
  &\multicolumn{2}{c|}{sorting by size,}
  &\multicolumn{2}{c|}{sorting by size,}
  &\multicolumn{2}{c|}{all techniques} \\
  &\multicolumn{2}{c|}{stream solve}&\multicolumn{2}{c|}{stream solve}
  &\multicolumn{2}{c|}{stream solve}&\multicolumn{2}{c|}{of section \ref{charac}} \\ 
   \cline {2-9} 
  &    solve&  subst.&     solve&  subst. &    solve&  subst. &  solve & subst.  \\ \hline
 3&       .02      &      .01      &       .01      &      .00      &       .00      &       .00      &      .01     &    0  \\
 4&       .14      &      .01      &       .15      &      .02      &       .02      &       .02      &
                                                                                                             .03      &    0  \\
 5&      1.6 \ \   &      .17      &      1.3 \ \   &      .23      &       .20      &       .16      &
                                                                                                             .08      &    0  \\
 6&     20.8 \ \   &     1.46      &     15.3 \ \   &     2.4 \ \   &      1.7 \ \   &      1.3 \ \   &
                                                                                                             .26      &    0  \\
 7&    206 \ \ \ \ &     9.0 \ \   &     85 \ \ \ \ &     9.0 \ \   &     16.3 \ \   &     15.4 \ \   &
                                                                                                             .77      &    0  \\
 8&  2,740 \ \ \ \ &    82.5 \ \   &    808 \ \ \ \ &    92.8 \ \   &    241 \ \ \ \ &    223 \ \ \ \ &
                                                                                                            2.7 \ \   &   .01 \\
 9& 53,940 \ \ \ \ & 1,190 \ \ \ \ & 13,335 \ \ \ \ & 1,482 \ \ \ \ &  4,645 \ \ \ \ &  2,675 \ \ \ \ &
                                                                                                            9.1 \ \   &   .01 \\
10&                &               &                &               & 60,140 \ \ \ \ & 20,360 \ \ \ \ &
                                                                                                           30 \ \ \ \ &   .03 \\
11&                &               &                &               &                &                &
                                                                                                           96 \ \ \ \ &   .07 \\
12&                &               &                &               &                &                &
                                                                                                          302 \ \ \ \ &   .13 \\
13&                &               &                &               &                &                &
                                                                                                          927 \ \ \ \ &   .34 \\
14&                &               &                &               &                &                &  2284 \ \ \ \ &   .42 \\
15&                &               &                &               &                &                &  7587 \ \ \ \ &  1.58 \\
16&                &               &                &               &                &                & 27970 \ \ \ \ &  3.16 \\ \hline
  \end{tabular}   
  \end{center}
  \caption{Times in sec for solving the symmetry conditions 
           (\ref{symcon}) for each degree $n$ by different methods 
           and for substituting the computed values into the symmetry 
           ansatz (\ref{symm}).}
  \label{table2}
\end{table*}

The impact of measures described in the previous section is 
shown in table \ref{table2}. 
The entries in columns with the header 'solve' give the time
in sec to solve the linear algebraic system resulting from splitting
the symmetry condition (\ref{symcon}) (denoted in the following by
$D_{[t,\tau]}(u,v)=0$) wrt.\ monomials in
$u,v,u^{-1},v^{-1}$. Columns with the header 'subst.' show
the time to substitute the solution into the symmetry ansatz
(\ref{symm}).\footnote{Columns (A)-(C) have been run on one 
Opteron 2.2 GHz core of a 32 core node (8 sockets x 4 cores per socket) with 128 GB memory.
In column (D) 
$n=3..15$ were run on one Xeon 2.4 GHz core of a 16 core node with 128 GB memory and 
$n=16$ was run on a single core of a machine equipped with 4 AMD Opteron
processors each having 12 cores running at 2.2.GHz and with 256 GB of main
memory.  Computation times turned out to be strongly dependent on the load on
the remaining cores of the node and the other nodes of the cluster.  The times in
column (D) are conservative times which are always reproducible. Sometimes
computations have been up to 30\% faster than shown in column (D).}

As described in the previous section, column (A) was produced with the 
original stream solver \cite{streamsol}. Sorting equations by size 
results in a speedup of a factor up to 4 shown in column (B) and column 
(C) shows the times when 1-term equations are applied first as described 
in section \ref{apply}. Computations in columns (A)-(C) use standard 
substitutions and in column (C) the explicit formulation of 1-term equations 
and complete splittings of equations. 
The drastic improvement shown in column (D) 
became possible through direct detection of vanishing variables (specialized 
partial splitting), repeated partial splitting, 
assigning nil to the value cell of a variable and thus avoiding substitutions
and enabling fast pruning of expressions from zero variables.

\begin{table*}[ht]
  \tabcolsep=3pt
  \begin{center}
  \begin{tabular}{|l|c|c|c|c|} \hline
                                   &      (E)           &       (F)           &    (G)     &   (H)     \\
                                  &$D_{[t,\tau]}(u,v)=0$&$D_{[t,\tau]}(u,v)=0$&$D_\tau I=0$& Iteration \\
                                   & at once            & in two stages       & first      & first     \\ \hline
formulation of ansatz for $D_\tau$ &    47.7            &    47.7             &  47.7      &   47.7    \\ \hline
multiple runs to find vanishing    &     $-$            &     $-$             &   $-$      &   777     \\
variables (see section \ref{iter}) &                    &                     &            &           \\ \hline
formulation and extraction of      &                    &                     &            &           \\
vanishing variables once from      &     $-$            &     $-$             &  238.7     &   $-$     \\
necessary condition  (\ref{nc1})   &                    &                     &            &           \\ \hline
formulation and complete           &                    &                     &            &           \\
splitting of $D_{[t,\tau]}u=0$     &    
                                         1624           &   
                                                             1624             &  621.3     &   4.5     \\ \hline
extraction of vanishing            &                    &                     &            &           \\
variables from complete set of     &     $-$            &    101.7            &   64.5     &  0.28     \\ 
conditions $D_{[t,\tau]}u=0$ once  &                    &                     &            &           \\ \hline
formulation and complete           &                    &                     &            &           \\
splitting of $D_{[t,\tau]}v=0$     &   
                                        1274            &    46.9             &   46.7     &  1.77     \\ \hline
complete solution of all           &                    &                     &            &           \\
remaining equations                &     365            &    125              &   142      &  92.7     \\ \hline
substitution of solution           &                    &                     &            &           \\
in $D_\tau(u,v)$                   &     2.95           &     0.39            &   0.35     &  0.18     \\ \hline
total time (rounded)               &    3314            &   1946              &  1161      &  924      \\ \hline

  \end{tabular}     
  \end{center}
  \caption{Times in sec of the whole symmetry computation for $n=13$}
  \label{table3}
\end{table*}

Table \ref{table3} compares four runs of the $n=13$ case.\footnote{All 
  times in this table include CPU time and garbage collection time of CSL 
  {\sc Reduce} running on one of 16 cores (4 sockets x 4 cores per socket) 
  Xeon \@ 2.4 GHz node with 128 GB memory.}  
In all these computations the program {\sc LSSS} is used to
solve the linear algebraic systems (the row 'complete solution of all
remaining equations'). {\sc LSSS} repeatedly applies 1-term equations
and sorts the remaining equations by size before starting the stream-solver.
What table \ref{table3} shows is the benefit of 
generating and solving the linear algebraic system in stages.
With the availability of {\sc LSSS} as an effective solver of large 
selection systems, the main cost shifted to the formulation of the linear
system.

In the first column the whole system is
formulated and solved at once. In the second column at first $D_{[t,\tau]}u=0$
is formulated, then 1-term equations are extracted and utilized to simplify
the ansatz for $D_\tau u$ and $D_\tau v$ before $D_{[t,\tau]}v=0$ is formulated and
the complete system is solved. This provides a speedup of nearly 1.7 for
$n=13$.

The third column gives the times when the whole computation is done in
3 stages.  In the first step the necessary condition 
(\ref{nc1}): $D_\tau I = \sum_{k=-k_0}^{k_0} a_k I^k$ is
formulated, and once vanishing variables are extracted
which are used to simplify the symmetry ansatz (\ref{symm}) and speed up
the formulation of the system.
Although that first step is an
additional computation not performed in the first two runs costing an
extra 238 sec, this is more than compensated afterwards by the speedup
of computing and solving $D_{[t,\tau]}(u,v)=0$ (also in two stages)
leading to another overall speedup factor of about 1.7.

The fourth column reports the times when at first 
a sequence of runs is performed where vanishing variables are extracted
and set to nil as explained in section \ref{iter}.
This leads to another modest overall speedup of 1.25 but the main advantage of
the computation in the fourth column is to lower the maximum memory
requirements that occur when the first half of symmetry conditions is formulated.

The next section describes how differences between two versions of the
computer algebra system {\sc Reduce} become important for large computations.
\begin{table*}[ht]
  \begin{center}
  \tabcolsep=3pt
  \begin{tabular}{|r|r|r|r|r|} \hline 
 n &  (I)  &  (J)  &   (K)   &   (L)   \\   
   &s solve&s tools&ns solve &ns tools \\ \hline
 3 & .0036 & .024  &  .035   &    .027 \\
 4 & .11   & .089  &  .11    &    .085 \\
 5 & .38   & .20   &  .3     &    .23  \\
 6 & 1.1   & .86   &  1.2    &    .80  \\
 7 & 4.3   & 2.97  &  4.9    &   3.75  \\
 8 & 20.4  & 13    &  24     &  16.9   \\
 9 & 128   & 73    &  162    &  81.3   \\
10 & 792   & 262   &  929    &   294   \\
11 & 7560  & 871   & 8615    &   972   \\
12 & DNF   & 3527  &   DNF   &  3232   \\
13 &       & 11786 &         & 11840   \\ \hline
  \end{tabular}
  \end{center}
  \caption{{\sc Maple} 14 times}
  \label{table2m1}
\vspace*{-20pt}
\begin{tabbing}
 \ \ \ \  ns tools: \= solving the union of both systems with the {\tt SolveTools:-Linear} command  \kill \\
 \ \ \ \  n         \> degree of the symmetry \\
 \ \ \ \  s solve:  \> solving the system (\ref{symcon}) with the
                       {\tt solve} command \\
 \ \ \ \  s tools:  \> solving the system (\ref{symcon}) with the
                       {\tt SolveTools:-Linear} command \\
 \ \ \ \  ns solve: \> solving both systems (\ref{nc}) and (\ref{symcon})
                       with the {\tt solve} command \\
 \ \ \ \  ns tools: \> solving both systems (\ref{nc}) and (\ref{symcon})
                        with the {\tt SolveTools:-Linear} command \\
 \ \ \ \  DNF:      \> does not finish in one week 
\end{tabbing}
\end{table*}

\section{{\sc Reduce} issues} \label{reduce}
The open-source computer algebra system {\sc Reduce} can be run on two
different LISP implementations.  They are PSL (Portable Standard Lisp) and CSL
(Codemist Standard Lisp). 

The computations of this article require a very large number of identifiers.
CSL has no restriction on the number of identifiers that can be used, whereas
PSL is usually limited to 65,000 identifiers. To run the symmetry computations
reported in this paper in PSL, an extended version of PSL {\sc Reduce} had
been developed by Winfried Neun that allows for 20 Mio identifiers. This
extended version of PSL can be downloaded for free \cite{reduce20m}.

When applied straightforwardly, PSL {\sc Reduce} typically runs somewhat
faster than CSL {\sc Reduce}.\footnote{The timings reflect the state of CSL/PSL
  at the time of writing the paper.  The performance of both LISP systems is
  under active development.}
However, CSL allows for compilation of critical routines
into C. When this is done, CSL runs in most cases as fast as PSL or even
faster. Compilation into C can be achieved by creating a package and making a
profiling run followed by a rebuild of the CSL-Reduce system.

A big advantage of CSL is its garbage collector. CSL uses both a copying and a
mark-sweep collector. Copying garbage collectors have little overhead but can
give only half of the memory to the application as the other half is needed
for copying. A mark-sweep garbage collector allows the use of all of the
memory by the application.  In CSL the copying collector is used as long as
the memory pressure is low, but switches to mark-sweep when the requirement
for more memory rises. This way, CSL leverages performance and available
memory. This feature of CSL made it possible to ultimately compute symmetries
of degree 15 with 115 GB of memory and symmetries of degree 16 on a different
machine with 256 GB.
\begin{table*}[ht]
  \begin{center}
  \tabcolsep=3pt
  \begin{tabular}{|r|c|c|c|c|c|c|} \hline 
n   &   (M)   &   (N)   &  (O)  &  (P)   &  (Q): read       & (M/(O+P+Q)) \\
    & unsimp. & unsorted& sorted&simplify&simplified sys.&factor of speedup \\ \hline
3   & .024    & .0012   & .0014 &  0     &  .010      & 2.1 \\
4   & .089    & .0084   & .0084 &  .01   &  .012      & 2.9 \\
5   & .20     & .011    & .0092 &  .01   &  .012      & 6.4 \\
6   & .86     & .036    & .037  &  .03   &  .016      & 10  \\
7   & 3.0     & .047    & .047  &  .09   &  .021      & 19  \\
8   & 13      & .19     & .22   &  .31   &  .037      & 23  \\
9   & 73      & .41     & .38   & 1.04   &  .060      & 49  \\
10  & 262     & 2.6     & 2.7   & 3.44   &  .118      & 42  \\
11  & 871     & 4.7     & 4.9   & 11.2   &  .228      & 53  \\
12  & 3527    & 32      & 31    & 31.7   &  .443      & 56  \\
13  & 11786   & 78      & 71    & 87.5   &  .940      & 74  \\
14  &         & 402     & 387   &        & 1.574      &     \\ \hline
  \end{tabular}    
  \end{center}
  \caption{{\sc Maple} 14 times for pre-simplified systems}
  \label{table3c}
\end{table*}

\begin{table*}[ht]
  \begin{center}
  \tabcolsep=3pt
  \begin{tabular}{|r|r|r|r|r|r|} \hline 
   & {\sc Maple} 14                           
   & \multicolumn{3}{c|}{{\sc LinBox}} 
   & {\sc Reduce} \\ \cline{2-6}
 n & SolveTools    
   & default & \multicolumn{2}{c|}{sparse} 
   & LSSS \\ \cline{2-6}
   & eqn.\ (\ref{symcon}) 
   & eqn.\ (\ref{symcon}) & eqn.\ (\ref{symcon}),(\ref{nc})  & eqn.\ (\ref{symcon}) 
   & eqn.\ (\ref{symcon}),(\ref{nc}) \\ \hline
 3 &     .024 &             &              &              &  .01 \\ 
 4 &     .09  &   .02       &     .02      &     .02      &  .03 \\ 
 5 &     .20  &             &     .13      &     .12      &  .08 \\ 
 6 &     .86  & 30.6\, \    &     .90      &    1.1\, \   &  .26 \\ 
 7 &    3     &             &   12.9\, \   &   14.9\, \   &  .77 \\ 
 8 &   13     &3080 \ \ \ \ &  210 \ \ \ \ &  283.5\, \   &  2.7 \\ 
 9 &   73     &             & 1812 \ \ \ \ & 2318 \ \ \ \ &  9.1 \\ 
10 &  262     &             &21210 \ \ \ \ &21610 \ \ \ \ &   30 \\ 
11 &  871     &             &              &              &   96 \\ 
12 &  3527    &             &              &              &  302 \\ 
13 &  11786   &             &              &              &  927 \\ 
14 &          &             &              &              & 2284 \\ 
15 &          &             &              &              & 7587 \\
16 &          &             &              &              &27970 \\  \hline
  \end{tabular}    
  \end{center}
  \caption{A comparison of times of {\sc Maple, LinBox} and {\sc Reduce}}
\label{tablexx}
\end{table*}

\section{Comparison with {\sc Maple}} \label{maple}
In this section we report on applying the computer algebra system {\sc Maple} 14 on
solving our linear systems. Table \ref{table2m1} shows
times\footnote{Computations were run on one Xeon 2.4 GHz core of a 16 core 
node with 128 GB memory} for solving the
symmetry conditions (\ref{symcon}) with the {\sc Maple} standard {\tt solve} command
in column (I) and with the {\tt SolveTools:-Linear} command in column (J) and
the times for solving the combined systems of additional conditions (\ref{nc})
and symmetry conditions (\ref{symcon}) in columns (K) and (L). 

We see that for small systems, {\tt solve} and {\tt SolveTools:-Linear} take
similar times. For larger systems ($N=11$) {\tt SolveTools:-Linear} is 9 times
as fast and even bigger systems could not be solve with the command 
{\tt solve} within a week.

Another observation is that neither {\tt solve} nor {\tt SolveTools:-Linear}
can take advantage of additional necessary equations except for $n=12$ in
column (L) to a small extent.
The best times for {\sc Maple} 14 in column (J) of table \ref{table2m1} and {\sc LSSS} ({\sc Reduce})
in column (D) in table \ref{table2} have been measured on the same nodes of
the same cluster and can be compared. To explain the 11786/927=12.7 times faster
performance of {\sc LSSS} for symmetries of degree 13 the following test has been made.
Table \ref{table3c} shows the times in sec for {\sc Maple} 14 procedure {\tt SolveTools:-Linear} to solve
\begin{itemize}
  \item the unsimplified system (\ref{symcon}) in column (M) (=column (J) of
        table \ref{table2m1}), 
  \item the same system simplified after 1-term equations have been solved
        repeatedly in column (N), 
  \item and then in addition all equations being sorted by size, shortest first in column (O).
\end{itemize}
The cost of these simplifications using the {\sc Reduce} procedures 
{\sc FindZeros} to find and utilize all 1-term equations repeatedly and {\sc LengthSort}
to sort the remaining system by size is shown in column (P). The time for 
{\sc Maple 14} to read the simplified system is given in column (Q). The rightmost
column gives the factor of speedup of {\sc Maple} if it would use at first the 
{\sc Reduce} procedures to simplify the system before solving it.
The potential speedup of {\sc Maple} is impressive and increasing with 
increasing size of the system.

\section{Comparison with {\sc LinBox}}  \label{linbox}
As described on its web page \cite{libo} 
{\sc LinBox} is a C++ template library for exact, high-performance
linear algebra computations with dense, sparse, and structured matrices
over the integers and over finite fields.

Because the solution of our linear system (\ref{symcon}) contains free
parameters, {\sc LinBox} can not be applied straightforwardly to solve our
system.  However, what can be computed easily is the rank of these
systems. Table \ref{tablexx} compares the best times of {\sc Maple} solving
the system (\ref{symcon}) (column (J) in table \ref{table2m1}) with the best
times of {\sc Reduce} achieved when solving (\ref{symcon}) and using 
additional necessary conditions (\ref{nc}) and with {\sc LinBox} at least
determining the dimension of the nullspace of (\ref{symcon})
and of (\ref{symcon}) together with (\ref{nc}). The timings of
{\sc LinBox} include reading of the sparse coefficient matrix of the linear
system which is the only way to enter data into LinBox.  

\section{The procedures} \label{procedures}
The following four procedures are available for the formulation and solution of
linear algebraic system with an emphasis on selection systems which have many
zero variables in their solution.

The procedure {\sc LSSS} can solve linear systems with an arbitrary number
of equations and unknowns, i.e. the linear system can be under- or
overdetermined. In case the linear system is inconsistent a message will be
printed. The unknowns to be solved for must be ordered ahead of the rest of
the variables. The coefficients of the linear variables can be rational
functions of parameters.

The procedure {\sc FindZeros} repeatedly picks all 1-term equations from
an input list of equations and sets the value cell of the vanishing variables
to nil. It returns a list of vanishing variables and list of remaining
equations.

The procedure {\sc LengthSort} effectively sorts a list of equations
by their size.

The procedure {\sc PruneZeros} drops all terms of the input expression
which should be zero due to an earlier run of {\sc LSSS} or {\sc FindZeros}.

The procedures {\sc LSSS, FindZero} and {\sc LengthSort} are freely available from 
\cite{liso} where more details about syntax are given.  These procedures
are also included as module in the open source computer algebra system {\sc
  Reduce}.  The server at \cite{liso} contains also files for all linear
systems for symmetries of degree 3 to 15 in {\sc Reduce} format and in
{\sc LinBox} format which can also be read by {\sc Maple}.

\section{Summary} \label{summary}
A class of sparse linear systems that allows efficient solution
strategies and that occurs frequently, for example, in integrability
investigations, is characterized by the vanishing of many of the
unknowns in its solution. In association with the purpose of such
systems we call them selection systems. A series of such systems that
we investigated results from symmetry investigations of a non-abelian
ODE system of Kontsevich. It is demonstrated how the vanishing of many
variables of a selection system can not only be used to speed up the
solution of the system but also to avoid formulating most of the
system and to apply the solution of the system to large
expressions. It is demonstrated how a pre-processor step in which
repeatedly 1-term equations are identified and applied and afterwards
the remaining equations are sorted by size could speed up {\sc Maple}
routines that are specialized in sparse systems by a considerable
factor.

\section*{Acknowledgement}
The first author would like to thank Winfried Neun for providing a PSL
version for large numbers of identifiers.  The second author would
like to thank Arthur C. Norman for very helpful discussions concerning
CSL.  Computations were run on computer hardware of the Sharcnet
consortium ({\tt www.sharcnet.ca}).

\end{document}